\documentclass{aip-cp}

\usepackage[numbers]{natbib}
\usepackage{rotating}
\usepackage{graphicx}

\begin{document}

% Title portion
\title{Studies on the Reduction of Radon Plate-Out}

\author[aff1]{M. Bruemmer}
\author[aff1]{M. Nakib}
\author[aff1]{R. Calkins}
\author[aff1,cor1]{J. Cooley}
\author[aff1]{S. Sekula}

\affil[aff1]{Department of Physics, Southern Methodist University, Dallas, TX 75275, USA}

\corresp[cor1]{Corresponding author: cooley@physics.smu.edu}

\newcommand{\Rn}[1]{\ensuremath{^{#1}\mathrm{Rn}}}
\newcommand{\Pb}[1]{\ensuremath{^{#1}\mathrm{Pb}}}
\newcommand{\Th}[1]{\ensuremath{^{#1}\mathrm{Th}}}
\def\in{\ensuremath{\mathrm{in}}}
\def\cm{\ensuremath{\mathrm{cm}}}
\def\h{\ensuremath{\mathrm{h}}}

\maketitle

\begin{abstract}
The decay of common radioactive gases, such as radon, produces stable isotopes by a sequence of daughter particles with varied half-lives. These daughter particles are a significant source of gamma, neutron, and alpha ($\alpha$) particle backgrounds that can mimic desired signals in dark matter and neutrinoless double beta decay experiments. In the LUMINA Laboratory at Southern Methodist University (SMU), studies of radon plate-out onto copper samples are conducted using one of XIA's first five UltraLo 1800 alpha counters. We present results from investigations into various mitigation approaches. A custom-built copper holder (in either plastic or metal) has been designed and produced to maximize the copper's exposure to $\Rn{220}$. The $\Rn{220}$ source is a collection of camping lantern mantles. We present the current status of control and experimental methods for addressing radon exposure levels.
\end{abstract}

% Head 1
\section{INTRODUCTION}
$\Pb{210}$ is a long-lived daughter of the decay of $\Rn{222}$.  The decay of $\Pb{210}$ to $\Pb{206}$ produces $\alpha$ and beta particles that are a primary background for dark matter experiments.  In this experiment, we use $\Rn{220}$ daughters (produced by a $\Th{232}$ source) as a proxy for $\Rn{222}$. $\Rn{220}$ has several advantages.  It is more short-lived than $\Rn{222}$ allowing for brief experiments and quick interpretations, and it is more easily procured than $\Rn{222}$. The daughters of radon decay are positively charged about 90\% of the time (c.f. \cite{kotrappa1981}), which allows them to plate-out on material surfaces but could also be used to prevent them from doing so by exposing them to an external electric field.

\section{THE EQUIPMENT}

For each exposure, three copper plates measuring $4 \: \in \times 4\: \in$ ($10 \: \cm \times 10\: \cm$) are placed into either a 3D-printed plastic or a machined stainless steel sample holder, each of identical geometric design.   The samples and their holder are then placed into a pressure cooker that had been modified to be an air-tight exposure chamber.  The exposure chamber is purged with nitrogen boil-off gas for 15 minutes before each exposure. Exposures each last five days. The exposure source is composed of eight camping lanterns mantles; such mantles manufactured prior to the 1990s contain $\Th{232}$. We labeled our mantles and always arranged them in the same locations and orientations on the floor of the exposure chamber. This ensured a similar exposure source configuration for every experiment and mitigated the fact that the exact radon emission rate of each mantle is not known. 

After exposure, the samples are counted in an XIA Ultra-Lo 1800 $\alpha$ particle counter \cite{ultralo2013, warburton2004, Gordon2009, McNally:2014eka}.  The XIA counter is a specialized ionization counter comprised of an active volume filled with argon, a lower grounded electrode that is a conductive tray holding the sample, and an upper pair of positively charged electrodes. Of these two electrodes, the anode sits directly above the sample, while the guard electrode surrounds and encloses the anode. Both electrodes are connected to charge-integrating preamplifiers whose output signals are digitized and then processed by a digital pulse shape analyzer. The background of the SMU XIA counter is $0.001 \alpha/(\cm^{2} \cdot \h)$.

\section{REDUCTION OF RADON PLATE-OUT USING ELECTRIC FIELDS (``E-SHIELD'')}

Prior to assembling the experimental equipment for this part of the study, we estimated the stopping potential required to bring to rest every $\Rn{220}$ daughter particle assuming that the parent Radon atom
is at rest when it decays and that, in the worst-case scenario, the daughter isotope gets all available kinetic energy from the two-body $\alpha$ decay of the parent $\Rn{220}$ 
(this is the ``non-thermalized'' (NT) case). Under those assumptions,
% Unnumbered equation
 \[
  {v_{NT} = \sqrt{\frac{2K_{daughter}}{m_{daughter}}}} \approx  3 \times 10^{5} \: \mathrm{m/s}.
\]
where $K_{daughter}$ ($m_{daughter}$) is the kinetic energy (mass) of the daughter isotope. We use the simplifying assumption that the electric field involved in stopping the daughter is uniform. Under this hypothesis, the electric field required to bring a particle at this speed to rest within the typical distance available inside the pressure cooker ($d \simeq 3\cm$) is then
\[ 
{E_{NT} = \frac{m_{daughter}a_{daughter}}{2e}} \approx 2 \times 10^{6} \: \mathrm{N/C}.
\]
where $a_{daughter}$ is the acceleration of the daughter required to bring it to zero velocity. We can then compute the potential required to generate this electric field:
\[
{V_{NT} = E_{NT}d \approx 60 \: \mathrm{kV}}.
\]
At standard temperature and pressure ($P$), and absent the $60\mathrm{kV}$ potential, the mean-free path of such a non-thermalized radon daughter is approximately
\[
{l = \frac{K_{daughter}}{\sqrt{2}\pi D_{daughter}P} \approx 1.5 \:\mathrm{m}}
\]
where $D_{daughter}$ is the cross-sectional area of the daughter ion. We expect the mean-free path to be reduced in the presence of an external electric field. Within the cost constraints of this project, we were unable to procure a direct current (DC) power supply with this maximum potential; instead, we employed a variable DC power supply capable of $5-35\mathrm{kV}$. Due to the short distances involved inside the pressure cooker, we were only able to operate the power supply at $6\mathrm{kV}$ due to dielectric breakdown of the air in the vessel. We expected that even at this potential, some measurable fraction of the radon daughters could be stopped before plate-out on the copper samples.

The experimental setup for this study includes the exposure chamber, the plastic sample holder, and the copper samples. In addition, the DC power supply is wired to a nickel-copper fabric anode cap that is slid over the top of the sample holder and shields the samples on 5 sides (the bottom is open). The anode is maintained at $6\mathrm{kV}$ and the body of the pressure cooker is grounded. Based on this geometry, we produced a simplified model of the electric field to understand how field lines connected would connect from the anode to ground. This map confirmed that positively charged radon daughters should be accelerated away from the copper samples toward the walls of the exposure vessel.

Three copper samples were placed into the sample holder and covered with the anode cap.  The exposure cycle was begun as described earlier. The first exposure was a control with all components present but no electric field, to isolate the effect of covering the copper with the anode cap.  The second exposure employed the  electric field.  Results from these tests are shown in Fig. \ref{fig:results} and Table \ref{tab:results}, and are discussed below.

\section{REDUCTION OF RADON PLATE-OUT USING NITROGEN PURGE}

Purging a volume with nitrogen boil-off is a commonly used technique to mitigate radon plate-out.  We conducted a second set of experiments aimed to compare the reduction in plate-out from the use of the electric field to the use of a nitrogen purge. For this set of experiments, the metal sample holder was employed and no anode cap was present, alllowing for greater exposure of the copper samples during the control phase.

Three copper samples were placed in identical positions in the holder as in the E-Shield experiment.  The exposure chamber volume was measured to be $0.22 \mathrm{ft}^{3}$.  For the control, the chamber was purged with nitrogen gas but no flow was otherwise employed. For the active experiment, we selected a  flow rate of 3 SCFH such that the nitrogen atmosphere of the chamber was replaced every five minutes.  Nitrogen gas entered the chamber at the top of the exposure chamber and exited through a flow meter and regulator at the bottom of the chamber.  Results of this study are shown in Fig. \ref{fig:results} and Table \ref{tab:results}.

\section{RESULTS AND SUMMARY}

\begin{figure}[h]
  \centerline{
    \includegraphics[width=0.45\linewidth]{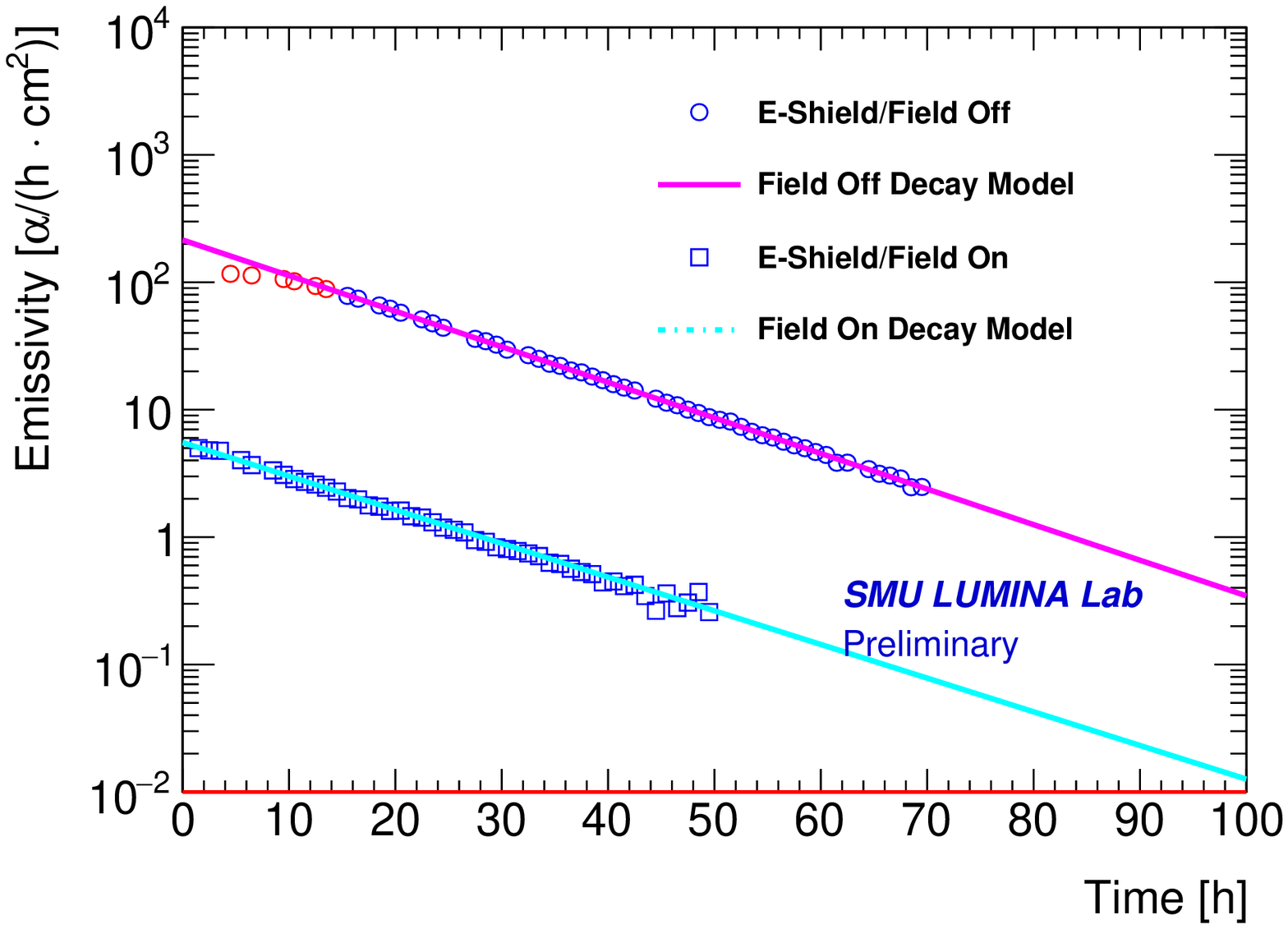}
    \includegraphics[width=0.45\linewidth]{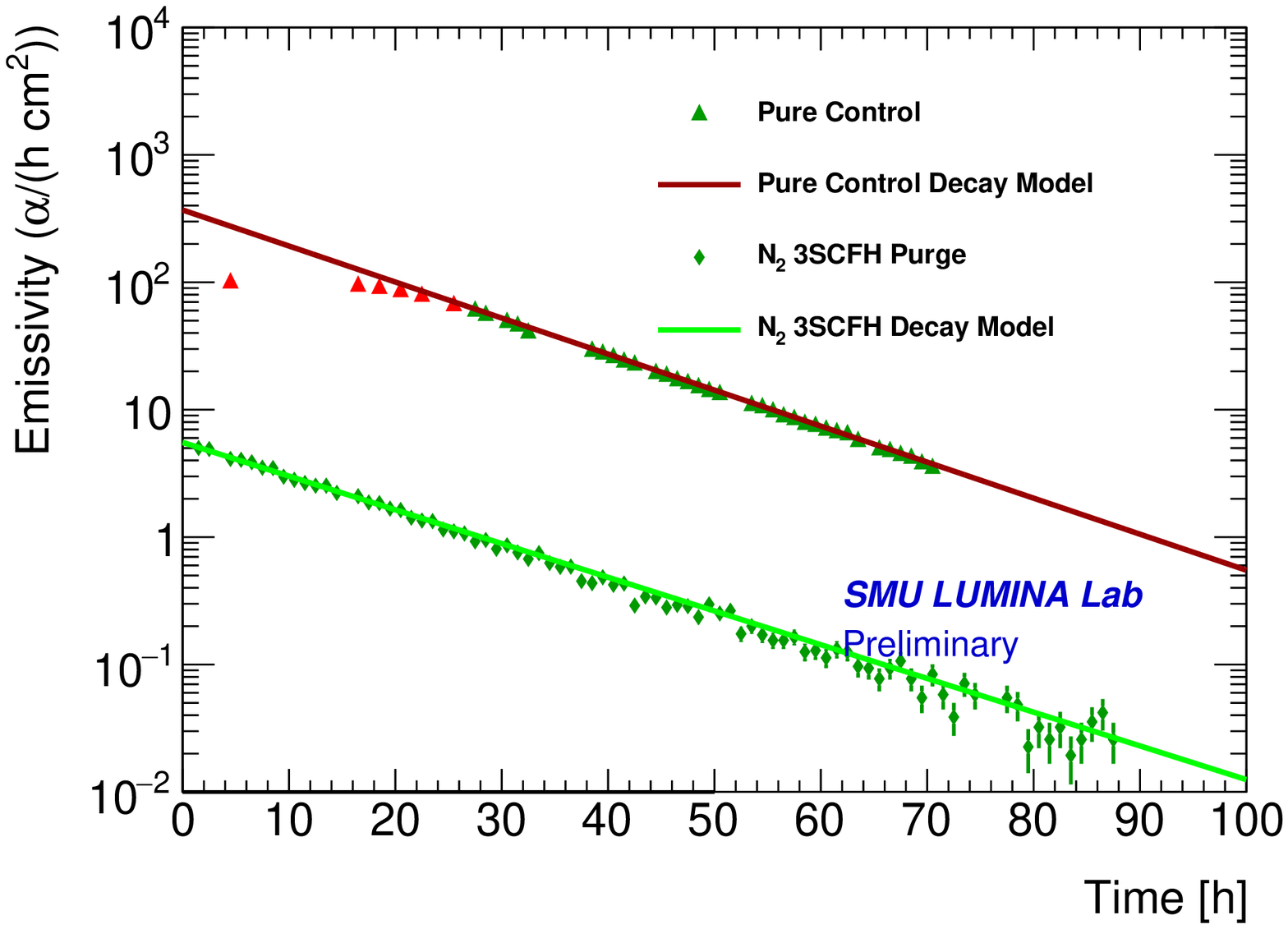}
  }
\caption{\label{fig:results}Emissivity vs. time for the E-Shield (left) and nitrogen purge (right) experiments. Shown are the emissivity data from the respective control experiments and active experiments. The smooth lines are the result of fitting an exponential decay model to the good-quality data. The red-colored data points from early counting times correspond to data from poor-quality periods when the $\alpha$ rate was high enough to saturate the XIA counter for our laboratory configuration of the equipment. All other data points are considered to be from periods of reliable counting.}
\end{figure}

The results of the studies are compared in Fig. \ref{fig:results} and Table \ref{tab:results}. In both the E-Shield and nitrogen purge experiments, we observe a significant reduction in the radon plate-out (at the level of 98\% or better). Future directions in these studies will be to improve the design of the radon exposure vessel to allow for the largest available potential to be achieved with our existing equipment ($35\mathrm{kV}$). We will study quantatively the relationship between voltage and radon plate-out, and nitrogen purge rate and radon plate-out. In the latter study, the future goal is a cost-benefit analysis of purge rate vs. outcome. 

% Table
\begin{table}[h]
\caption{Comparison of the results from various exposures.  The control experiment always involves exposure to the radon source with no mitigation technique employed.  The E-shield, field-off experiment included the anode cap but no electric field. }
\label{tab:results}
\tabcolsep7pt\begin{tabular}{lc}
\hline
\tch{1}{c}{t}{Experiment}  & \tch{1}{c}{t}{Emissivity\\  \lbrack $\alpha/(\cm^{2} \cdot \h$)\rbrack}   \\
\hline
control & 26.8 $\pm$ 0.3 \\
3 SCFH Ni Purge & 0.42 $\pm$ 0.04 \\
\hline\hline
E-shield, field-off & 15.9 $\pm$ 0.2  \\
E-shield, field-on & 0.45 $\pm$ 0.04 \\
\hline
\end{tabular}
\end{table}

% Acknowledgement
\section{ACKNOWLEDGMENTS}
This material is based upon work supported by the National Science Foundation under Grant Number (1151869), the SMU Hamilton Scholar program and SMU Engaged Learning.   Any opinions, findings, and conclusions or recommendations expressed in this material are those of the author(s) and do not necessarily reflect the views of the National Science Foundation.  We would also like to thank Kevin Cieszowski, John Cotton, Tim Mulone, Lacey Porter, Hang Qiu, and Randall Scalise for their contributions to this project.

% References

\nocite{*}
\bibliographystyle{aipnum-cp}%
\bibliography{2015_bruemmer}%

\end{document}